\documentclass[usegraphicx]{mn2e}

\title[The effect of ISM turbulence on the gravitational instability of
       galactic discs]
      {The effect of ISM turbulence on the gravitational instability of
       galactic discs}

\author[V. Hoffmann and A. B. Romeo]
       {Volker Hoffmann$^{1}$\thanks{E-mail: volker@physik.uzh.ch}
        and Alessandro B. Romeo$^{2}$\\
        $^{1}$Institute for Theoretical Physics,
              University of Z\"{u}rich,
              CH-8057 Z\"{u}rich, Switzerland\\
        $^{2}$Onsala Space Observatory,
              Chalmers University of Technology,
              SE-43992 Onsala, Sweden}

\begin{document}

\date{Accepted 2012 July 5.
      Received 2012 July 5; in original form 2012 March 6}

\pagerange{\pageref{firstpage}--\pageref{lastpage}}

\pubyear{2012}

\maketitle

\label{firstpage}

\begin{abstract}
We investigate the gravitational instability of galactic discs, treating
stars and cold interstellar gas as two distinct components, and taking into
account the phenomenology of turbulence in the interstellar medium (ISM),
i.e.\ the Larson-type scaling relations observed in the molecular and atomic
gas.  Besides deriving general properties of such systems, we analyse a large
sample of galaxies from The H\,\textsc{i} Nearby Galaxy Survey (THINGS), and
show in detail how interstellar turbulence affects disc instability in
star-forming spirals.  We find that turbulence has a significant effect on
both the inner and the outer regions of the disc.  In particular, it drives
the inner gas disc to a regime of transition between two instability phases
and makes the outer disc more prone to star-dominated instabilities.
\end{abstract}


\section{INTRODUCTION}

Today, 30 years after the pioneering work by Larson (1981), observations and
simulations of the interstellar medium (ISM) are revealing its turbulent
nature with higher and higher fidelity (see, e.g., Elmegreen \& Scalo 2004;
McKee \& Ostriker 2007; Romeo et al.\ 2010).  A fundamental aspect of ISM
turbulence is the existence of scaling relations between the mass column
density ($\Sigma$), the 1D velocity dispersion ($\sigma$), and the size of
the region over which such quantities are measured ($\ell$):
\begin{equation}
\Sigma\propto\ell^{a},\;\;\;
\sigma\propto\ell^{b}.
\end{equation}
The values of $a$ and $b$, and the range of scales spanned by $\ell$ depend
on which ISM component we consider.  In this paper we focus on cold
interstellar gas, which is highly supersonic and hence strongly compressible,
and which is known to play an important role in the gravitational instability
of galactic discs (e.g., Lin \& Shu 1966; Jog \& Solomon 1984a,\,b; Bertin \&
Romeo 1988, and references therein).

In the molecular gas, $\mathrm{H}_{2}$, the scaling exponents are $a\approx0$
and $b\approx\frac{1}{2}$, and Eq.\ (1) holds up to scales of a few 100 pc.
In fact, both Galactic and extragalactic giant molecular clouds (GMCs) are
fairly well described by Larson's scaling laws, $\Sigma=constant$ and
$\sigma\propto\ell^{1/2}$, although the uncertainties are still large (e.g.,
Larson 1981; Solomon et al.\ 1987; Bolatto et al.\ 2008; Heyer et al.\ 2009;
Hughes et al.\ 2010; Kauffmann et al.\ 2010; Lombardi et al.\ 2010;
S\'{a}nchez et al.\ 2010; Azimlu \& Fich 2011; Ballesteros-Paredes et
al.\ 2011; Field et al.\ 2011; Kritsuk \& Norman 2011; Roman-Duval et
al.\ 2011; Beaumont et al.\ 2012).  Besides, Larson-type scaling relations
have now been observed, for the first time, in the dense star-forming clumps
of a high-redshift galaxy (Swinbank et al.\ 2011).

In the atomic gas, H\,\textsc{i}, the scaling exponents are instead
$a\sim\frac{1}{3}$ and $b\sim\frac{1}{3}$, and Eq.\ (1) seems to hold up to
scales of a few kpc.  A Kolmogorov scaling for both $\sigma$ and $\Sigma$ is
suggested by the observed power spectra of H\,\textsc{i} intensity
fluctuations, and is also consistent with other measurements (e.g., Lazarian
\& Pogosyan 2000; Elmegreen et al.\ 2001; Begum et al.\ 2006; Kim et
al.\ 2007; Dutta et al.\ 2008; Roy et al.\ 2008; Dutta et al.\ 2009a,\,b;
Block et al.\ 2010; Bournaud et al.\ 2010; Dutta et al.\ 2010; Dutta 2011;
Combes et al.\ 2012; Zhang et al.\ 2012).  Note, however, that the
uncertainties are larger than in the $\mathrm{H}_{2}$ case.  For example,
high-resolution simulations of supersonic turbulence suggest a Burgers
scaling for both $\sigma$ and $\Sigma$, i.e.\ $a\sim\frac{1}{2}$ and
$b\sim\frac{1}{2}$ (e.g., Fleck 1996; Kowal \& Lazarian 2007; Kowal et
al.\ 2007; Kritsuk et al.\ 2007; Schmidt et al.\ 2008; Price \& Federrath
2010).  Other recent simulation surveys suggest that the scaling exponent $a$
is significantly affected by turbulence forcing (Federrath et al.\ 2009,
2010) and self-gravity (Collins et al.\ 2012).

In spite of such a burst of interest in ISM turbulence, and in spite of the
dynamical importance of cold interstellar gas, there have been very few
theoretical works aimed at evaluating the effect of turbulence on disc
instability.  In fact, traditional stability analyses do not take into
account the scale-dependence of $\sigma$ (or $\Sigma$), but identify $\sigma$
with the typical 1D velocity dispersion observed at galactic scales.  The
first theoretical work devoted to the gravitational instability of turbulent
gas discs was made by Elmegreen (1996), who assumed Larson-type scaling
relations [see Eq.\ (1)] and investigated the case $a=-1$ and
$b=\frac{1}{2}$.  He found that the disc is always stable at large scales and
unstable at small scales.  Romeo et al.\ (2010) also assumed Larson-type
scaling relations, but explored the whole range of values for $a$ and $b$.
They showed that turbulence has an important effect on the gravitational
instability of the disc: it excites a rich variety of stability regimes,
several of which have no classical counterpart.  See in particular the
`\emph{stability map of turbulence}' (fig.\ 1 of Romeo et al.\ 2010), which
illustrates such stability regimes and populates them with observations,
simulations and models of interstellar turbulence.

In the gravitational instability of galactic discs, there is an important
interplay between stars and cold interstellar gas (e.g., Agertz et al.\ 2009;
Elmegreen 2011; Forbes et al.\ 2011; Cacciato et al.\ 2012).  The
gravitational coupling between these two components does not alter the form
of the local stability criterion, $Q_{\mathrm{eff}}\geq1$, but makes the
effective $Q$ parameter different from both the stellar and the gaseous
Toomre (1964) parameters (Bertin \& Romeo 1988; Romeo 1992, 1994; Elmegreen
1995; Jog 1996; Rafikov 2001; Shen \& Lou 2003; Elmegreen 2011; Romeo \&
Wiegert 2011).  The gravitational coupling between stars and gas also changes
the least stable wavelength (Jog 1996), among other diagnostics.

What is the effect of ISM turbulence in this more realistic context?  The
first published attempt to answer this question was made by Shadmehri \&
Khajenabi (2012).  They considered two-component discs of stars and turbulent
gas, chose $a$ and $b$ so as to sample five of the seven stability regimes
found by Romeo et al.\ (2010), and studied the dispersion relation
numerically.  Their study suggests that turbulence has a significant effect
on disc instability even when stars are taken into account.  The goal of our
paper is to answer the question above in detail, extending previous work
along two directions:
\begin{enumerate}
\item We perform a rigorous stability analysis of two-com\-ponent turbulent
  discs, motivated by observations of ISM turbulence in nearby galaxies.  In
  particular, we consider two complementary cases: H\,\textsc{i} plus
  $\mathrm{H}_{2}$, and gas plus stars.  In the first case, we examine the
  dispersion relation analytically, and illustrate how the gravitational
  coupling between H\,\textsc{i} and $\mathrm{H}_{2}$ modifies the main
  stability regimes of gas turbulence, which were originally derived
  neglecting such a coupling (Romeo et al.\ 2010).  In the second case, we
  show that there are four stability regimes of galactic interest, similar to
  those analysed above, but in only one of them do stars play a
  non-negligible role.  We then focus on such a regime, and illustrate how
  gas turbulence affects the onset of gravitational instability in the disc,
  i.e.\ the local stability threshold and the corresponding characteristic
  wavelength.
\item We apply this analysis to a large sample of star-form\-ing spirals from
  The H\,\textsc{i} Nearby Galaxy Survey (THINGS), previously analysed by
  Leroy et al.\ (2008) and Romeo \& Wiegert (2011), and illustrate how ISM
  turbulence affects a full set of stability diagnostics: the condition for
  star-gas decoupling, the effective $Q$ parameter, and the least stable
  wavelength.
\end{enumerate}

The rest of the paper is organized as follows.  The (in)stability of
two-component turbulent discs is analysed in Sect.\ 2, our application to
THINGS spirals is shown in Sect.\ 3, the relation between our results and
those of Shadmehri \& Khajenabi (2012) is discussed in Sect.\ 4, and the
conclusions are drawn in Sect.\ 5.

\section{(IN)STABILITY OF TWO-COMPONENT TURBULENT DISCS}

\subsection{Summary of the one-component case}

Here we summarize some of the results found by Romeo et al.\ (2010), which
are fundamental to a proper understanding of Sects 2.2--2.4.

The dispersion relation of a turbulent and realistically thick gas disc is
\begin{equation}
\omega^{2}=\kappa^{2}-2\pi G\Sigma(k)\,k+\sigma^{2}(k)\,k^{2},
\end{equation}
where $\omega$ and $k$ are the frequency and the wavenumber of the
perturbation, and $\kappa$ is the epicyclic frequency.  $\Sigma(k)$ and
$\sigma(k)$ are the mass column density and the 1D velocity dispersion
measured over a region of size $\ell=1/k$, as inferred from observations
(see, e.g., Elmegreen \& Scalo 2004; McKee \& Ostriker 2007; Romeo et
al.\ 2010):
\begin{equation}
\Sigma(k)=\Sigma_{0}\left(\frac{k}{k_{0}}\right)^{-a},\;\;\;
\sigma(k)=\sigma_{0}\left(\frac{k}{k_{0}}\right)^{-b}.
\end{equation}
If the disc has volume density $\rho$ and scale height $h$, then
$\Sigma\approx2\rho\ell$ for $\ell\la h$ and $\Sigma\approx2\rho h$ for
$\ell\ga h$.  The range $\ell\la h$ corresponds to the case of 3D turbulence
(GMCs and H\,\textsc{i} at small scales), whereas the range $\ell\ga h$
corresponds to the case of 2D turbulence (H\,\textsc{i} at large scales).
The quantity $\ell_{0}=1/k_{0}$ introduced in Eq.\ (3) is the fiducial scale
at which $\Sigma$ and $\sigma$ are observed.  This is also the scale at which
the Toomre parameter $Q$ and other stability quantities are measured, so that
$Q_{0}=\kappa\sigma_{0}/\pi G\Sigma_{0}$.

The scaling exponents $a$ and $b$ have an important effect on the shape of
the dispersion relation [Eq.\ (2)], and hence on the condition for local
gravitational instability ($\omega^{2}<0$).  As $a$ and $b$ vary, turbulence
drives the disc across seven stability regimes, three of which are densely
populated by observations, simulations and models of galactic turbulence (see
fig.\ 1 of Romeo et al.\ 2010):
\begin{itemize}
\item For $b<\frac{1}{2}\,(1+a)$ and $-2<a<1$ (hereafter \emph{Regime A}),
  the stability of the disc is controlled by $Q_{0}$: the disc is stable at
  all scales if and only if $Q_{0}\geq\overline{Q}_{0}$, where
  $\overline{Q}_{0}$ depends on $a$, $b$ and $\ell_{0}$.  This is the domain
  of H\,\textsc{i} turbulence.  Both H\,\textsc{i} observations and
  high-resolution simulations of supersonic turbulence are consistent with
  the scaling $a=b$.  In such a case, the local stability criterion
  degenerates into $Q_{0}\geq1$, as if the disc were non-turbulent and
  infinitesimally thin.
\item For $b>\frac{1}{2}\,(1+a)$ and $-2<a<1$ (hereafter \emph{Regime C}),
  the stability of the disc is no longer controlled by $Q_{0}$: the disc is
  always unstable at small scales (i.e.\ as $k\rightarrow\infty$) and stable
  at large scales (i.e.\ as $k\rightarrow0$).
\item For $b=\frac{1}{2}\,(1+a)$ and $-2<a<1$ (hereafter \emph{Regime B}),
  the disc is in a phase of transition between stability \`{a} la Toomre
  (Regime A) and instability at small scales (Regime C).  This is the domain
  of $\mathrm{H}_{2}$ turbulence.  Note, however, that even small deviations
  from Larson's scaling laws can drive the disc into Regime A or Regime C,
  and thus have a strong impact on its gravitational instability.
\end{itemize}

\begin{figure}
\centering
\includegraphics[scale=1.]{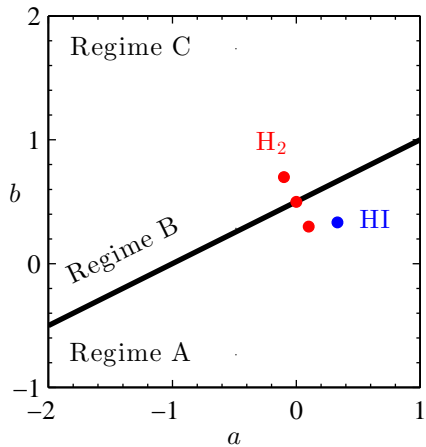}
\caption{The main stability regimes of one-component turbulent discs (see
  Sect.\ 2.1).  Also shown are the two-component cases illustrated in
  Fig.\ 2: H\,\textsc{i} in Regime A ($a_{1}=b_{1}=\frac{1}{3}$) plus
  $\mathrm{H}_{2}$ in Regime A ($a_{2}=0.1$, $b_{2}=0.3$), Regime B
  ($a_{2}=0.0$, $b_{2}=0.5$), or Regime C ($a_{2}=-0.1$, $b_{2}=0.7$).}
\end{figure}

Since Regimes A--C are fundamental to a proper understanding of Sects
2.2--2.4, we show them in Fig.\ 1.  Note, however, that this simple figure is
not meant to be a substitute for fig.\ 1 of Romeo et al.\ (2010), which
illustrates all seven stability regimes and their relation to the
phenomenology of ISM turbulence.

\subsection{Dispersion relation and general properties}

Until now we have considered H\,\textsc{i} and $\mathrm{H}_{2}$ separately.
How does the stability scenario change when H\,\textsc{i} and
$\mathrm{H}_{2}$ are considered together?  And how does it change when both
gas and stars are taken into account?  We will answer these questions here
and in Sects 2.3 and 2.4.

When H\,\textsc{i} and $\mathrm{H}_{2}$ are considered together, their
gravitational coupling changes how the disc responds to perturbations.  The
dispersion relation can be expressed in a form that is particularly useful
for discussing the stability properties of the disc:
\begin{equation}
\left(\omega^{2}-\mathcal{M}_{1}^{2}\right)
\left(\omega^{2}-\mathcal{M}_{2}^{2}\right)=
\left(\mathcal{P}_{1}^{2}-\mathcal{M}_{1}^{2}\right)
\left(\mathcal{P}_{2}^{2}-\mathcal{M}_{2}^{2}\right),
\end{equation}
where
\begin{equation}
\mathcal{M}_{i}^{2}\equiv
\kappa^{2}-2\pi G\Sigma_{i}(k)\,k+\sigma_{i}^{2}(k)\,k^{2},
\end{equation}
\begin{equation}
\mathcal{P}_{i}^{2}\equiv
\kappa^{2}+\sigma_{i}^{2}(k)\,k^{2},
\end{equation}
and $i=1,2$.%
\footnote{The dispersion relation of an $N$-component turbulent disc is
  $\sum_{i=1}^{N} (\mathcal{M}_{i}^{2}-\mathcal{P}_{i}^{2}) /
  (\omega^{2}-\mathcal{P}_{i}^{2}) = 1$, as can easily be inferred from
  eq.\ (22) of Rafikov (2001).  This equation cannot be expressed in a form
  similar to Eq.\ (4), and will not be used in the rest of the paper.}
Note that $\omega^{2}=\mathcal{M}_{i}^{2}(k)$ is the one-component dispersion
relation for potential-density waves [cf.\ Eq.\ (2)], while
$\mathcal{P}_{i}^{2}(k)$ describes sound waves modified by rotation (and
turbulence).  Since $\mathcal{M}_{i}^{2}(k)-\mathcal{P}_{i}^{2}(k)$
represents the self-gravity of component $i$, the right-hand side of Eq.\ (4)
measures the strength of gravitational coupling between the two components.

Eqs (4)--(6) are also applicable to two-component discs of gas and stars,
even though the stellar component is collisionless and non-turbulent.  This
is because stars can be accurately treated as a fluid when analysing the
stability of galactic discs (Bertin \& Romeo 1988; Rafikov 2001), and because
the equations above are valid whether each fluid is turbulent or not.
Remember, in fact, that the phenomenology of turbulence is encapsulated in
$\Sigma_{i}(k)$ and $\sigma_{i}(k)$ without altering the form of those
equations.  When the disc is made of gas (g) and stars ($\star$),
$\Sigma_{\mathrm{g}}(k)$ and $\sigma_{\mathrm{g}}(k)$ are given by Eq.\ (3),
while the stellar quantities are not.  $\Sigma_{\star}(k)$ is the reduced
surface density, $\Sigma_{\star}(k)=\Sigma_{\star}/(1+kh_{\star})$, where the
$k$-dependent factor results from the finite scale height of the stellar
layer (Vandervoort 1970; Romeo 1992, 1994; Elmegreen 2011).  In contrast,
$\sigma_{\star}$ is the radial velocity dispersion and does not depend on
$k$, since the pressure term in the dispersion relation is unaffected by disc
thickness (see again Vandervoort 1970).  The gaseous and stellar Toomre
parameters are then defined as
$Q_{\mathrm{g}0}=\kappa\sigma_{\mathrm{g}0}/\pi G\Sigma_{\mathrm{g}0}$ and
$Q_{\star}=\kappa\sigma_{\star}/\pi G\Sigma_{\star}$.

\begin{figure*}
\includegraphics[scale=.97]{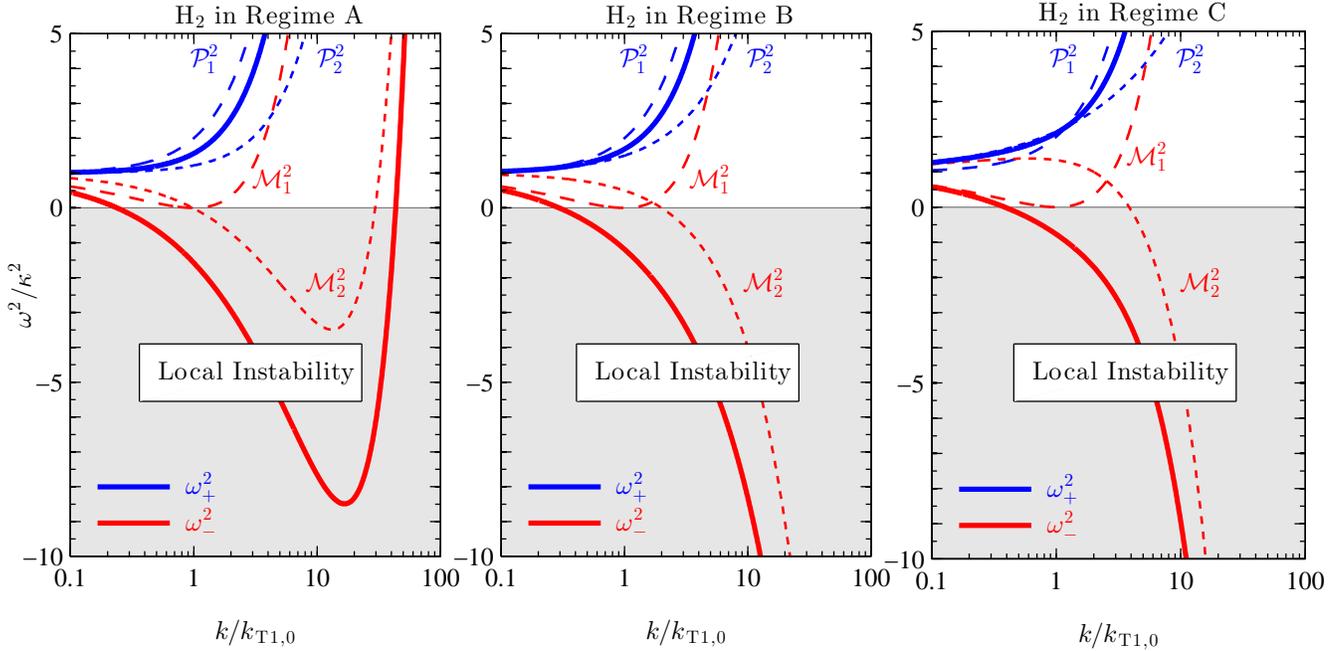}
\caption{The two branches of the dispersion relation, $\omega_{+}^{2}(k)$ and
  $\omega_{-}^{2}(k)$, vs.\ their one-component counterparts,
  $\mathcal{P}_{i}^{2}(k)$ and $\mathcal{M}_{i}^{2}(k)$, in stability regimes
  of galactic interest.  These quantities are measured in units of
  $\kappa^{2}$, the square of the epicyclic frequency, while $k$ is measured
  in units of $k_{\mathrm{T1,0}}=\kappa^{2}/2\pi G\Sigma_{1,0}$, the Toomre
  wavenumber of component $i=1$ at scale $\ell=\ell_{0}$.  The cases
  illustrated represent a disc made of marginally stable H\,\textsc{i} (in
  Regime A) and unstable $\mathrm{H}_{2}$ (in Regimes A--C).  The scaling
  exponents are specified in Fig.\ 1.  The other independent quantities are
  as follows: $k_{0}=8\,k_{\mathrm{T1,0}}$, $Q_{1,0}=1$,
  $\Sigma_{2,0}/\Sigma_{1,0}=1$ and $\sigma_{2,0}/\sigma_{1,0}=\frac{1}{2}$.
  These relations imply that $Q_{2,0}=\frac{1}{2}$ and
  $k_{0}=\frac{1}{2}\,k_{\mathrm{J2,0}}$, where $k_{\mathrm{J2,0}}=2\pi
  G\Sigma_{2,0}/\sigma_{2,0}^{2}$ is the Jeans wavenumber of component $i=2$
  ($\mathrm{H}_{2}$) at scale $\ell=\ell_{0}$.}
\end{figure*}

As Eq.\ (4) is quadratic in $\omega^{2}$, it can be solved with elementary
methods.  The discriminant is positive, so there are two real roots:
\begin{equation}
\omega_{\pm}^{2}=
\frac{1}{2}\left[\left(\mathcal{M}_{1}^{2}+\mathcal{M}_{2}^{2}\right)
\pm\sqrt{\Delta}\,\right],
\end{equation}
\begin{equation}
\Delta=
\left(\mathcal{M}_{1}^{2}-\mathcal{M}_{2}^{2}\right)^{2}+4
\left(\mathcal{P}_{1}^{2}-\mathcal{M}_{1}^{2}\right)
\left(\mathcal{P}_{2}^{2}-\mathcal{M}_{2}^{2}\right).
\end{equation}
This means that the dispersion relation has two branches that do not cross,
$\omega_{+}^{2}(k)\neq\omega_{-}^{2}(k)$, except possibly as $k\rightarrow0$
or $k\rightarrow\infty$.  The functions $\omega_{\pm}^{2}(k)$ satisfy two
basic properties, which constrain the gravitational instability of the disc
and generalize the stability constraints found in the classical two-component
case (Jog \& Solomon 1984a; Bertin \& Romeo 1988).  Such properties are
stated and proved below, and can easily be visualized with the help of
Fig.\ 2.  The cases illustrated represent a disc made of marginally stable
H\,\textsc{i} (in Regime A) and unstable $\mathrm{H}_{2}$ (in Regimes A--C).
\begin{itemize}
\item \emph{Property I:} $\omega_{-}^{2}(k)$ lies below both
  $\mathcal{M}_{1}^{2}(k)$ and $\mathcal{M}_{2}^{2}(k)$, i.e.\ a
  two-component self-gravitating disc is more unstable (or less stable) than
  each component, whether this is turbulent or not.  This can be proved by
  noting that $\Delta$ is larger than
  $(\mathcal{M}_{1}^{2}-\mathcal{M}_{2}^{2})^{2}$, so that
  $\sqrt{\Delta}>|\mathcal{M}_{1}^{2}-\mathcal{M}_{2}^{2}|$.  In turn, this
  implies that $\omega_{-}^{2}<\mathcal{M}_{\mathrm{min}}^{2}$, where
  $\mathcal{M}_{\mathrm{min}}^{2}$ is the smallest $\mathcal{M}_{i}^{2}$ for
  a given $k$.
\item \emph{Property II:} $\omega_{+}^{2}(k)$ is bounded by
  $\mathcal{P}_{1}^{2}(k)$ and $\mathcal{P}_{2}^{2}(k)$, i.e.\ this branch is
  always stable and represents sound waves modified by rotation (and
  turbulence).  To prove this, note that the inequality
  $\sqrt{\Delta}>|\mathcal{M}_{1}^{2}-\mathcal{M}_{2}^{2}|$ also implies that
  $\omega_{+}^{2}>\mathcal{M}_{\mathrm{max}}^{2}$, where
  $\mathcal{M}_{\mathrm{max}}^{2}$ is the largest $\mathcal{M}_{i}^{2}$ for a
  given $k$.  Note also that $\omega_{+}^{2}$ cannot be smaller than
  $\mathcal{P}_{\mathrm{min}}^{2}$ or larger than
  $\mathcal{P}_{\mathrm{max}}^{2}$, otherwise Eq.\ (4) would not hold.
  Therefore it must be $\mathcal{P}_{\mathrm{min}}^{2}\leq\omega_{+}^{2}\leq
  \mathcal{P}_{\mathrm{max}}^{2}$.
\end{itemize}

\subsection{\textmd{H\,\textsc{i}} plus $\mathrm{H}_{2}$}

In Sect.\ 2.1, we have summarized the main stability regimes of one-component
turbulent discs.  Let us now extend the discussion to two-component discs of
H\,\textsc{i} and $\mathrm{H}_{2}$, analysing three cases of galactic
interest (see again Fig.\ 2).

\subsubsection{$\mathrm{H}_{2}$ in Regime A}

The response of each component is driven by pressure at small scales and by
rotation at large scales, while self-gravity acts more strongly at
intermediate scales (see sect.\ 2.7 of Romeo et al.\ 2010).  This means that
the gravitational coupling between the two components is negligible as
$k\rightarrow0$ and $k\rightarrow\infty$, and so is the right-hand side of
Eq.\ (4).  Therefore the two branches of the dispersion relation behave
asymptotically as $\mathcal{M}_{1}^{2}(k)$ and $\mathcal{M}_{2}^{2}(k)$,
i.e.\ they converge to $\kappa^{2}$ as $k\rightarrow0$ and diverge positively
as $k\rightarrow\infty$.  Since the potentially unstable branch
$\omega_{-}^{2}(k)$ lies below $\mathcal{M}_{i}^{2}(k)$ (cf.\ Property I) and
$\mathcal{M}_{i}^{2}(k)$ has a minimum for $k>0$, $\omega_{-}^{2}(k)$ must
also have a global minimum below $\kappa^{2}$.  Thus the disc is stable \`{a}
la Toomre, like each component (see left panel of Fig.\ 2).

\subsubsection{$\mathrm{H}_{2}$ in Regime C}

The response of H\,\textsc{i} is similar to the previous case, while
$\mathrm{H}_{2}$ behaves differently (see sect.\ 2.5 of Romeo et al.\ 2010).
The self-gravity term gets dominant for large $k$ and makes
$\mathcal{M}_{\mathrm{H}2}^{2}(k)$ negative.  So $\omega_{-}^{2}(k)$ is also
negative in this limit (cf.\ Property I).  For small $k$,
$\mathcal{M}_{\mathrm{H}2}^{2}(k)$ is positive since it is dominated by the
pressure term ($b>1$) and/or the rotation term ($b\leq1$).  As neither
$\mathcal{M}_{\mathrm{H2}}^{2}(k)$ nor $\mathcal{M}_{\mathrm{HI}}^{2}(k)$ is
driven by self-gravity at large scales, the right-hand side of Eq.\ (4) is
negligible as $k\rightarrow0$.  So $\omega_{-}^{2}(k)$ is positive in this
limit, like $\mathcal{M}_{\mathrm{HI}}^{2}(k)$ and
$\mathcal{M}_{\mathrm{H2}}^{2}(k)$.  The disc is then unstable at small
scales and stable at large scales, like $\mathrm{H}_{2}$ itself (see right
panel of Fig.\ 2).

\subsubsection{$\mathrm{H}_{2}$ in Regime B}

The behaviour of $\mathrm{H}_{2}$ is intermediate between the previous two
cases (see sect.\ 2.3 of Romeo et al.\ 2010).  A similar flow of arguments
shows that the disc is in a phase of transition between stability \`{a} la
Toomre and instability at small scales, like $\mathrm{H}_{2}$ itself.  The
middle panel of Fig.\ 2 illustrates the phase of small-scale instability,
which occurs for $k_{0}\leq k_{\mathrm{J2,0}}=2\pi
G\Sigma_{\mathrm{H2,0}}/\sigma_{\mathrm{H2,0}}^{2}$ (see Hoffmann 2010 for a
detailed analysis).  Note how the two components contribute to the
gravitational instability of the disc, and how their coupling widens the
range of unstable scales.

\subsection{Gas plus stars}

This case involves three components: H\,\textsc{i}, $\mathrm{H}_{2}$ and
stars.  In nearby spiral galaxies, H\,\textsc{i} and $\mathrm{H}_{2}$ have
distinct domains: H\,\textsc{i} dominates the outer regions of the gas disc,
while $\mathrm{H}_{2}$ dominates the inner regions (e.g., Leroy et
al.\ 2008).  We can then consider H\,\textsc{i} and $\mathrm{H}_{2}$
separately.  This makes sense here because we already know how the
gravitational coupling between H\,\textsc{i} and $\mathrm{H}_{2}$ modifies
the main stability regimes of gas turbulence (see Sect.\ 2.3).  What we now
want to understand is the role that stars play in this stability scenario.
Let us then distinguish two cases:
\begin{enumerate}
\item \emph{Stars plus $\mathrm{H}_{2}$.}  Since the stellar component
  populates Regime A (like H\,\textsc{i}) and $\mathrm{H}_{2}$ populates
  Regimes A--C, this case is qualitatively similar to the set of cases
  analysed in Sect.\ 2.3.  So there are three stability regimes: stability
  \`{a} la Toomre, instability at small scales, and a phase of stability
  transition.  Note that such a variety of regimes is driven by
  $\mathrm{H}_{2}$ turbulence.  The stellar component can only modify the
  shape of the dispersion relation; it cannot change the type of stability
  regime.  Note also that there is a mismatch between two important scales.
  One is the characteristic scale of stellar instabilities,
  $L_{\star}=\sigma_{\star}^{2}/\pi G\Sigma_{\star}$, which is typically
  $\sim1\;\mbox{kpc}$ (see, e.g., Binney \& Tremaine 2008).  The other is the
  largest scale at which $\mathrm{H}_{2}$ turbulence has been observed,
  $L_{\mathrm{H2}}\sim100\;\mbox{pc}$ (e.g., Bolatto et al.\ 2008).  Since
  $L_{\star}$ is one order of magnitude larger than $L_{\mathrm{H2}}$, the
  stellar component cannot play a significant role in such stability regimes.
  Therefore this is essentially a one-component case, driven and dominated by
  $\mathrm{H}_{2}$.  In Sect.\ 3, we will show that such stability regimes
  can indeed be frequent in nearby star-forming spirals.
\item \emph{Stars plus \textup{H\,\textsc{i}}.}  As both components populate
  Regime A, this is a case of stability \`{a} la Toomre: $\omega_{-}^{2}(k)$
  has a global minimum, which determines whether the disc is stable for all
  wavenumbers or not (cf.\ Sect.\ 2.3.1).  In contrast to case (i),
  H\,\textsc{i} turbulence reaches scales as large as 1--10 kpc (e.g., Kim et
  al.\ 2007; Dutta 2011).  This makes it possible for the stellar component
  to `interact' with H\,\textsc{i} turbulence and contribute significantly to
  two-fluid instabilities, as in the classical case of stars plus
  non-turbulent gas.
\end{enumerate}

As discussed above, case (ii) represents the only stability regime in which
stars play a non-negligible role.  We then focus on this case, and analyse
how gas turbulence affects the onset of gravitational instability in the
disc, i.e.\ the local stability threshold and the corresponding
characteristic wavelength.  The effect of disc thickness is well known in
this context (Romeo 1992, 1994; Elmegreen 2011; Romeo \& Wiegert 2011).  So
we do not take that effect into account.

\subsubsection{The stability threshold}

As this is a Toomre-like case, the local stability criterion can be expressed
in the usual form $Q_{\mathrm{eff}}\geq1$, where $Q_{\mathrm{eff}}$ is the
effective $Q$ parameter.  In the classical case of stars plus non-turbulent
gas, $Q_{\mathrm{eff}}$ depends on three parameters: $Q_{\star}$,
$Q_{\mathrm{g}}$ and $s=\sigma_{\mathrm{g}}/\sigma_{\star}$.  For analysing
$Q_{\mathrm{eff}}$ in detail, it is useful to factor out the dependence on
$Q_{\star}$, $Q_{\mathrm{eff}}=Q_{\star}/\,\overline{Q}$, and study the
stability threshold $\overline{Q}$ as a function of $s$ and
$q=Q_{\mathrm{g}}/Q_{\star}$ (Romeo \& Wiegert 2011).  When gas turbulence is
taken into account via Eq.\ (3), $\overline{Q}$ depends on five parameters:
\begin{equation}
s_{0}\equiv\frac{\sigma_{\mathrm{g}0}}{\sigma_{\star}},\;\;\;
q_{0}\equiv\frac{Q_{\mathrm{g}0}}{Q_{\star}},
\end{equation}
$a$, $b$ and
\begin{equation}
\mathcal{L}_{0}\equiv\ell_{0}\,k_{\mathrm{T}\star},
\end{equation}
where $k_{\mathrm{T}\star}=\kappa^{2}/2\pi G\Sigma_{\star}$ is the stellar
Toomre wavenumber.  A general five-parameter study of $\overline{Q}$ is not
more useful than a targeted few-parameter analysis.  This is because $a$, $b$
and $\mathcal{L}_{0}$ are tightly constrained by observations, and because
their observed values fall within a single stability regime (remember that
this is a region of the parameter space where the disc has similar stability
properties).  For these reasons, we analyse $\overline{Q}$ as a function of
$s_{0}$ and $q_{0}$, choosing observationally motivated values of $a$, $b$
and $\mathcal{L}_{0}$: $a=b=\frac{1}{3}$, which is the typical scaling of
H\,\textsc{i} turbulence (see Sect.\ 1), and $\mathcal{L}_{0}=1.0\pm0.5$,
which are the median and $1\sigma$ scatter of $\mathcal{L}_{0}$ in the outer
discs of THINGS spirals (where H\,\textsc{i} dominates; see Sect.\ 3).  The
range $0.5\leq\mathcal{L}_{0}\leq1.5$ is also representative of clumpy
galaxies at intermediate and high redshifts.%
\footnote{Puech (2010) analysed two such galaxy samples at $z\approx0.6$ and
  $z\approx2$.  The median properties of the discs are summarized in his
  table 1 (see also his sect.\ 3.2).  Using those data, we find that the
  stellar Toomre wavenumber is
  $k_{\mathrm{T}\star}\approx0.3\;\mbox{kpc}^{-1}$ at $z\approx0.6$ and
  $k_{\mathrm{T}\star}\approx0.2\;\mbox{kpc}^{-1}$ at $z\approx2$.  The
  spatial resolution is $\ell_{0}\approx7\;\mbox{kpc}$ and
  $\ell_{0}\approx5\;\mbox{kpc}$ in the two cases (Puech, private
  communication).  This yields $\mathcal{L}_{0}\approx2$ at $z\approx0.6$ and
  $\mathcal{L}_{0}\approx1$ at $z\approx2$.  Thus, even at intermediate and
  high redshifts, $\mathcal{L}_{0}$ is remarkably close to unity and lies
  within the $1\sigma$ scatter computed from THINGS.}
%

\begin{figure}
\centering
\includegraphics[scale=1.]{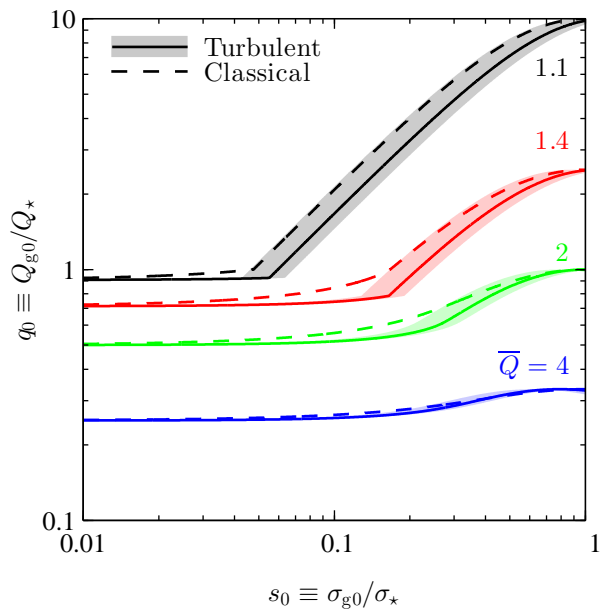}
\caption{Contour lines of the stability threshold,
  $\overline{Q}(s_{0},q_{0})=constant$, for discs of stars and turbulent
  H\,\textsc{i} vs.\ the classical case of stars plus non-turbulent gas.  The
  solid lines and the shaded regions correspond to the median and the
  $1\sigma$ scatter of $\mathcal{L}_{0}$ in the outer discs of THINGS
  spirals.}
\end{figure}

Fig.\ 3 shows a contour map of the stability threshold $\overline{Q}$ for
classical and turbulent discs.  Consider the classical case first, and look
at the contour levels $\overline{Q}=1.1$ and $\overline{Q}=1.4$.  Their slope
changes abruptly across the line $q_{0}=1$, showing that there are two
distinct stability regimes.  This fact has a simple explanation in terms of
star-gas decoupling (Bertin \& Romeo 1988; Romeo \& Wiegert 2011).  When
$s_{0}\la0.2$ and $q_{0}\sim1$, $\omega_{-}^{2}(k)$ has two minima: one at
small $k$, where the response of the stellar component peaks; and the other
at large $k$, where gas dominates.  For $q_{0}<1$, the gaseous minimum is
deeper than the stellar one, and therefore it controls the onset of disc
instability.  Vice versa, for $q_{0}>1$, it is the stellar minimum that
determines the stability threshold.  The line $q_{0}=1$ separates gas- from
star-dominated regimes even when $s_{0}\ga0.2$, but the transition is smooth
in this case since $\omega_{-}^{2}(k)$ has a single minimum.  In the
turbulent case, each contour level is on average shifted down.  As
$\overline{Q}$ increases in the same direction, this means that turbulence
lowers the stability threshold, i.e.\ it tends to stabilize the disc.  In
Sect.\ 3, we will evaluate the statistical significance of this effect.

\subsubsection{The characteristic wavelength}

The global minimum of $\omega_{-}^{2}(k)$ provides another useful stability
diagnostic: the least stable wavelength
$\lambda_{\mathrm{min}}=2\pi/k_{\mathrm{min}}$ (see Jog 1996 for the
classical case).  When the disc is marginally stable, the value of
$\lambda_{\mathrm{min}}$ is of particular interest.  It is the wavelength at
which instability first appears as $Q_{\star}$ drops below $\overline{Q}$.
This wavelength can be written as
$\overline{\lambda}=\overline{\Lambda}\,\lambda_{\mathrm{T}\star}$, where
$\lambda_{\mathrm{T}\star}=2\pi/k_{\mathrm{T}\star}$.  The characteristic
wavelength $\overline{\Lambda}$ depends on the same parameters as
$\overline{Q}$.  So we adopt the same approach as before, and analyse
$\overline{\Lambda}$ as a function of $s_{0}$ and $q_{0}$ for observationally
motivated values of $a$, $b$ and $\mathcal{L}_{0}$.

\begin{figure}
\centering
\includegraphics[scale=1.]{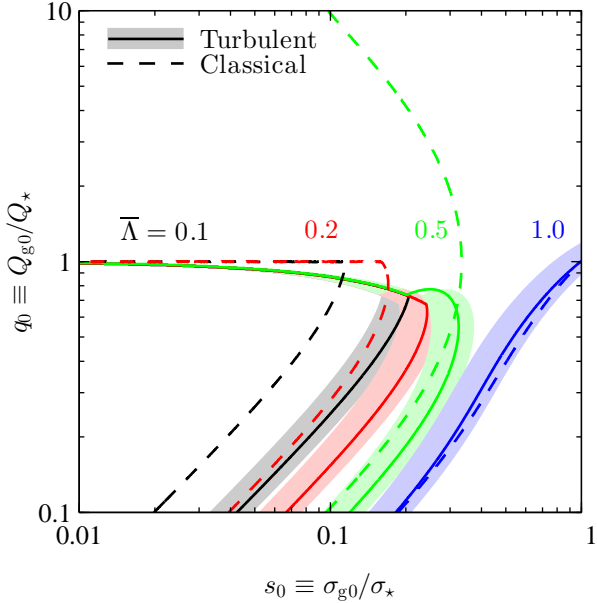}
\caption{Contour lines of the characteristic wavelength,
  $\overline{\Lambda}(s_{0},q_{0})=constant$, for discs of stars and
  turbulent H\,\textsc{i} vs.\ the classical case of stars plus non-turbulent
  gas.  The solid lines and the shaded regions correspond to the median and
  the $1\sigma$ scatter of $\mathcal{L}_{0}$ in the outer discs of THINGS
  spirals.}
\end{figure}

Fig.\ 4 shows a contour map of the characteristic wavelength
$\overline{\Lambda}$ for classical and turbulent discs.  In the classical
case, the contour levels $\overline{\Lambda}=0.1$ and
$\overline{\Lambda}=0.2$ are truncated above $q_{0}=1$.  This tells us that
such short characteristic wavelengths occur only when stars and gas are
decoupled and gas dominates.  In fact, star-dominated instabilities appear at
longer wavelengths: $\overline{\Lambda}\ga0.3$ (Bertin \& Romeo 1988).  Note
also that the contour $\overline{\Lambda}=0.5$ is a separatrix.  Levels below
0.5 are on the left of this curve (and connected to the transition line),
while levels above 0.5 are on the right.  In the turbulent case, each contour
level below 0.3 is on average shifted to the right, i.e.\ in the direction of
increasing $\overline{\Lambda}$.  This means that turbulence shortens the
characteristic wavelength when stars and gas are decoupled and gas dominates.
An opposite, although weaker, effect is detectable for
$\overline{\Lambda}\geq1$.  Other regimes are also affected, but in a more
complex way.  This is especially true for $\overline{\Lambda}\sim0.5$, since
the separatrix of the parameter plane shifts to larger values.  Last but not
least, note how turbulence bends the transition line down, favouring
star-dominated regimes.  In Sect.\ 3, we will analyse these effects in
detail.

\section{APPLICATION TO THINGS SPIRALS}

We now consider a sample of twelve nearby star-forming spirals from THINGS:
NGC 628, 3198, 3184, 4736, 3351, 6946, 3627, 5194, 3521, 2841, 5055, and
7331.  For these galaxies, a detailed analysis by Leroy et al.\ (2008)
provides high-quality measurements of kinematics, as well as stellar and
gaseous surface densities, at a constant spatial resolution of 800 pc.

Leroy et al.\ (2008) also analysed the stability of those galaxies, treating
the ISM as a single non-turbulent component, gravitationally coupled to
stars, with surface density
$\Sigma_{\mathrm{g}}=\Sigma_{\mathrm{HI}}+\Sigma_{\mathrm{H2}}$ and velocity
dispersion $\sigma_{\mathrm{g}}=11\;\mbox{km\,s}^{-1}$.  Such a value of
$\sigma_{\mathrm{g}}$ fits the H\,\textsc{i} data well, but is twice as large
as the typical $\mathrm{H}_{2}$ velocity dispersion observed in nearby spiral
galaxies (Wilson et al.\ 2011).  To represent both H\,\textsc{i} and
$\mathrm{H}_{2}$ well, we choose $\sigma_{\mathrm{g}}=8\;\mbox{km\,s}^{-1}$.
This value lies within the $1\sigma$ scatter of $\sigma_{\mathrm{HI}}$
($11\pm3\;\mbox{km\,s}^{-1}$; Leroy et al.\ 2008) and $\sigma_{\mathrm{H2}}$
($6.1\pm2.9\;\mbox{km\,s}^{-1}$; Wilson et al.\ 2011), and therefore allows
us to carry out an unbiased stability analysis of THINGS spirals.

The constant spatial resolution of 800 pc used by Leroy et al.\ (2008) makes
their data particularly appropriate for analysing the effect of H\,\textsc{i}
turbulence at galactic scales.  H\,\textsc{i} dominates the gas surface
density in the outer disc, typically for $R>0.43\;R_{25}$, where $R_{25}$ is
the optical radius (Leroy et al.\ 2008).  We then treat gas as turbulent for
$R>0.43\;R_{25}$, and assume Larson-type scaling relations [see Eq.\ (3)]
with $\ell_{0}=800\;\mbox{pc}$,
$\Sigma_{\mathrm{g}0}=\Sigma_{\mathrm{g}0}(R)$ as tabulated by Leroy et
al.\ (2008), and $\sigma_{\mathrm{g}0}=8\;\mbox{km\,s}^{-1}$ (see above).
Concerning $a$ and $b$, we analyse the case $a=b=\frac{1}{3}$ in detail,
since it represents H\,\textsc{i} observations fairly well (see Sect.\ 1).
We have also studied the case $a=b=\frac{1}{2}$, as representative of
high-resolution simulations of supersonic turbulence (see Sect.\ 1), but here
we will only mention it when discussing the results of our stability
analysis.  Hereafter we will refer to the model described above as
\emph{Model 1}.

\subsection{The condition for star-gas decoupling}

\begin{figure*}
\includegraphics[scale=.98]{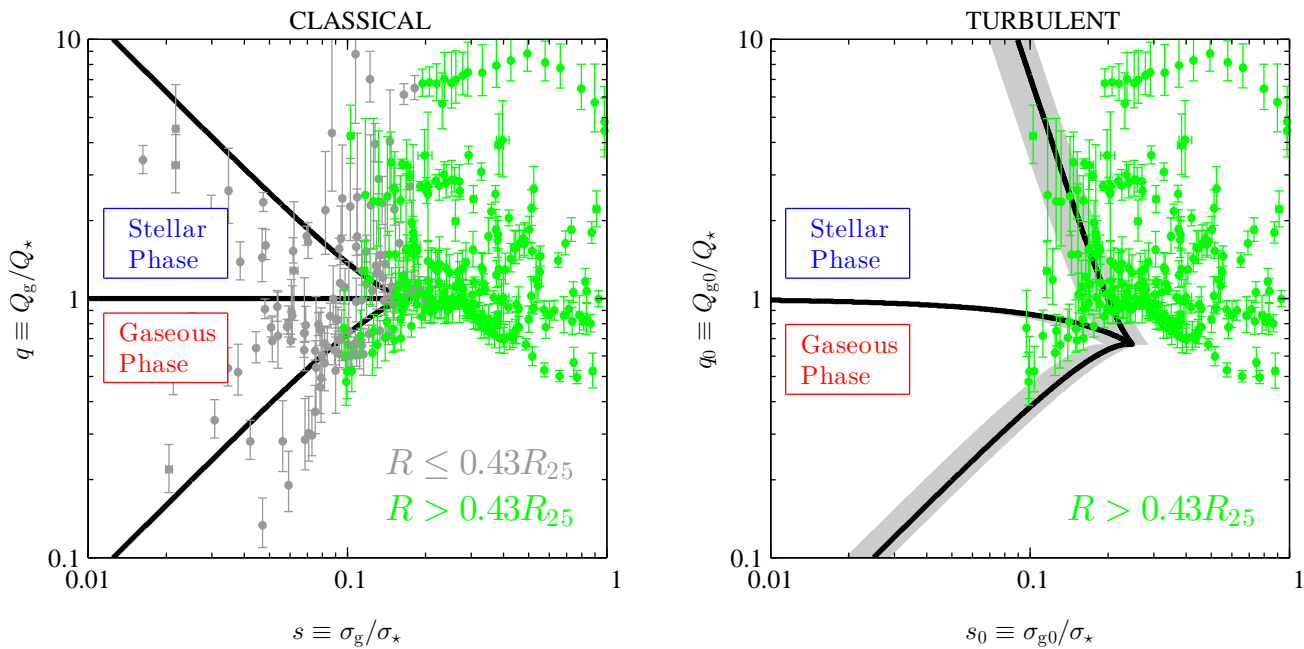}
\caption{The parameter plane populated by THINGS spirals (Model 1), and the
  `two-phase region' where stars and gas contribute separately to the
  gravitational instability of the disc: neglecting gas turbulence (left),
  and taking into account H\,\textsc{i} turbulence for $R>0.43\;R_{25}$
  (right).  In the turbulent case, the solid lines and the shaded regions
  correspond to the median and the $1\sigma$ scatter of $\mathcal{L}_{0}$ in
  that radial range.  Data from THINGS are coloured according to the dominant
  component: $\mathrm{H}_{2}$ for $R\leq0.43\;R_{25}$, and H\,\textsc{i} for
  $R>0.43\;R_{25}$.}
\end{figure*}

In Sect. 2.4, we have seen that there is a region in the parameter plane where $\omega_-^2(k)$ has two minima. This is the {\textquoteleft}two-phase region{\textquoteright} introduced by Bertin \& Romeo (1988) and further investigated by Romeo \& Wiegert (2011).

Fig.\ 5 shows the two-phase region for classical and turbulent discs. Within this region, stars and gas are dynamically decoupled and the disc is susceptible to instabilities at two different wavelengths, where the responses of the two components peak. In the stellar phase the disc is more susceptible to long-wavelength instabilities, whereas in the gaseous phase it is dominated by short-wavelength instabilities. Along the transition line between the phases, neither component dominates and instabilities occur both at short and at long wavelengths. Outside the two-phase region, the two components are strongly coupled and instabilities occur at intermediate wavelengths.

We populate the parameter plane with measurements taken from the sample of spiral galaxies, and colour-code them by radius. We draw the turbulent two-phase region corresponding to the median and $1\sigma$ scatter of $\mathcal{L}_0$ for $R > 0.43 \; R_{25}$. Note the following points:

\begin{enumerate}

  \item The two-phase region of a classical disc is symmetric about $q = 1$. This symmetry is broken for a turbulent disc because gas (dominant for $q < 1$) follows turbulent scaling, but stars (dominant for $q > 1$) do not.

  \item The turbulent two-phase region is larger than the classical one. This follows from the fact that turbulence pushes the minima of $\omega_-^2(k)$ further apart, and the maximum between them further up, so as to favour star-gas decoupling.  

  \item The transition line appears unaffected by the scatter of $\mathcal{L}_0$. This is because the shape of the two-phase region depends on $s$ and $q$, and $q$ is not affected by turbulence ($q=q_0$) if $a=b$.

  \item Turbulence increases the size of the stellar phase more than that of the gaseous phase. Recall that the boundary of the two-phase region is marked by the disappearance of the non-dominant peak, i.e. the gas peak in the stellar phase and vice versa. Since turbulence affects the gaseous peak more than the stellar peak, the size of the stellar phase is affected more than that of the gaseous phase. For $R > 0.43 \; R_{25}$, this causes a significant number of measurements to populate the stellar phase.

  \item For $R \leq 0.43 \; R_{25}$, we find that $f_2 = 61\%$ of all points populate the two-phase region, two-thirds of them in the gaseous phase. In such cases, the onset of gravitational instability is controlled by H$_2$. Turbulence is expected to play an important role in this process at scales smaller than about $100 \; \mathrm{pc}$ (see Sect. 2.4). For $R > 0.43 \; R_{25}$, only $4\%$ of all points populate this region for a classical disc. This fraction increases to $22\%$ for a turbulent disc with $a = b = \frac{1}{3}$, and to $52\%$ for $a=b=\frac{1}{2}$.

\end{enumerate}

\subsection{The effective $Q$ parameter}

\begin{figure*}
\includegraphics[scale=.99]{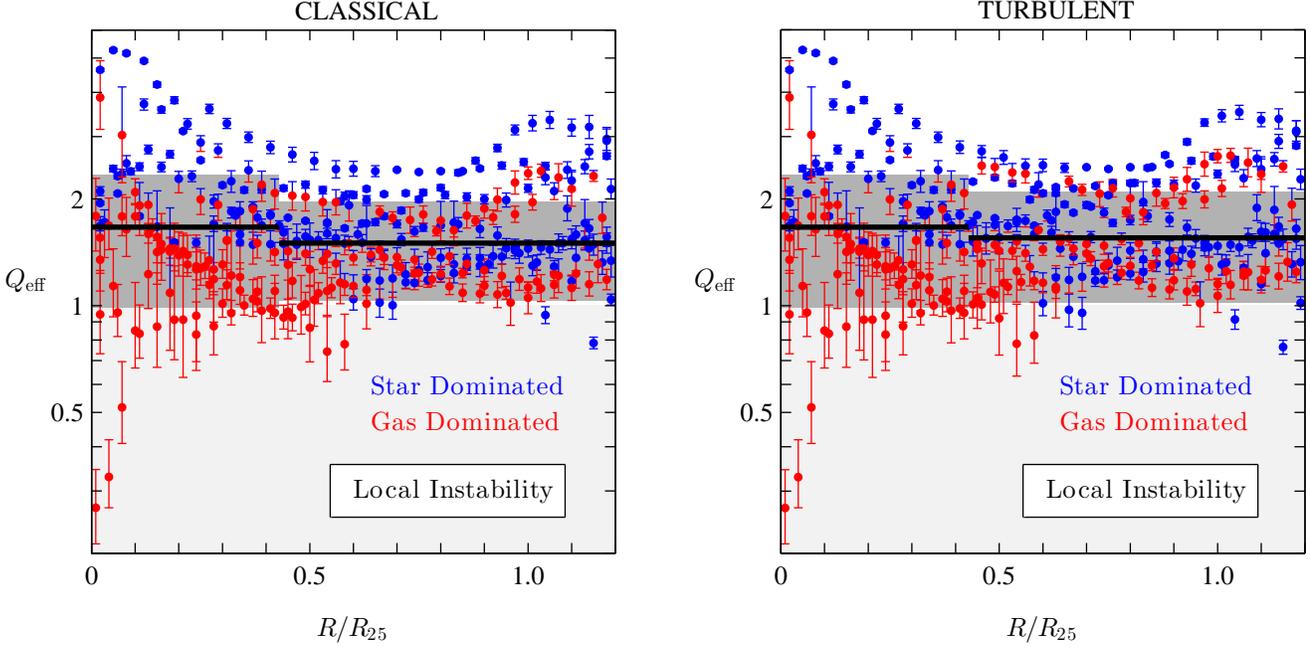}
\caption{Radial profile of the effective $Q$ parameter,
  $Q_{\mathrm{eff}}(R)$, for THINGS spirals (Model 1): neglecting gas
  turbulence (left), and taking into account H\,\textsc{i} turbulence for
  $R>0.43\;R_{25}$ (right).  The disc is locally unstable for
  $Q_{\mathrm{eff}}<1$ (light grey shading).  For each measurement, we
  indicate whether the stability level is dominated by the stellar
  ($Q_{\star}<Q_{\mathrm{g}0}$) or the gaseous ($Q_{\mathrm{g}0}<Q_{\star}$)
  component.  Thick black lines and dark grey shading indicate the median and
  $1\sigma$ scatter of $Q_{\mathrm{eff}}$ in the two radial ranges.}
\end{figure*}

Fig.\ 6 shows radial profiles of the effective $Q$ parameter, $Q_{\mathrm{eff}} = Q_{\mathrm{eff}}(R)$, for our sample of galaxies. In the left panel, we neglect gas turbulence. On the right, we consider turbulent H\,\textsc{i} ($a=b=\frac{1}{3}$) for $R > 0.43 \; R_{25}$. Values of $Q_{\mathrm{eff}}$ smaller than unity mean gravitational instability. We indicate the median and $1\sigma$ scatter of $Q_{\mathrm{eff}}$ for radii smaller and larger than $R = 0.43 \; R_{25}$. We also colour-code the component that contributes more to disc instability according to the classical condition: gas for $Q_{\mathrm{g}0} < Q_\star$, and stars for $Q_\star < Q_{\mathrm{g}0}$ (Romeo \& Wiegert 2011).

For $R \leq 0.43 \; R_{25}$, $Q_{\mathrm{eff}}$ spans a wide range of values, with $13\%$ of points in the unstable regime. Here $56\%$ of points are gas-dominated and tend to be less stable than the star-dominated points (the median value of $Q_\mathrm{eff}$ is $Q_{\mathrm{eff,g}} \approx 1.3$ and $Q_{\mathrm{eff},\star} \approx 2.3$ in the two cases). For $R > 0.43 \; R_{25}$, the range spanned by $Q_{\mathrm{eff}}$ is tighter and only $4\%$ of measurements are in the unstable regime. Here the majority ($61\%$) of points are star-dominated, and there is no clear difference in $Q_{\mathrm{eff}}$ between star- and gas-dominated points ($Q_{\mathrm{eff,g}} \approx 1.3$ and $Q_{\mathrm{eff},\star} \approx 1.7$).

Introducing turbulent scaling for $R > 0.43 \; R_{25}$ only has a small effect on the measurements. For $a = b = \frac{1}{3}$, the median of $Q_{\mathrm{eff}}$ increases by $3\%$ and the $1\sigma$ scatter by $15\%$. For $a = b = \frac{1}{2}$, the median increases by $6\%$ and the $1\sigma$ scatter by $26\%$. This suggests that turbulence tends to stabilize the disc (the median increases), although the magnitude of this effect is small and depends on the non-turbulent value of $Q_{\mathrm{eff}}$ (the scatter increases).

The stabilizing effect of turbulence seems at odds with results from Romeo et al. (2010), who found that the stability of gaseous discs is unaffected by turbulence if $a = b$. The difference lies, of course, in the gravitational coupling of stars and gas. Consider the approximation for the effective $Q$ parameter introduced by Romeo \& Wiegert (2011):

\begin{equation}
\frac{1}{Q_{\mathrm{eff}}}=
\left\{\begin{array}{ll}
       {\displaystyle\frac{W}{Q_{\star}}+\frac{1}{Q_{\mathrm{g}}}}
                       & \mbox{if\ }Q_{\star}\geq Q_{\mathrm{g}}\,, \\
                       &                                            \\
       {\displaystyle\frac{1}{Q_{\star}}+\frac{W}{Q_{\mathrm{g}}}}
                       & \mbox{if\ }Q_{\mathrm{g}}\geq Q_{\star}\,,
       \end{array}
\right.
\end{equation}
\begin{equation}
W=
\frac{2\sigma_{\star}\sigma_{\mathrm{g}}}
     {\sigma_{\star}^{2}+\sigma_{\mathrm{g}}^{2}}\,.
\end{equation}
We see that, even if $Q_{\mathrm{g}} = Q_{\mathrm{g}0}$, the scaling $\sigma_{\mathrm{g}}  = \sigma_{\mathrm{g}0} ( \ell / \ell_0 )^{b}$ affects the weight factor $W(\sigma_\star,\sigma_{\mathrm{g}})$. The strength of this effect is determined by the power-law slope $b$. Therefore the effective $Q$ parameter of turbulent discs always differs from the classical case.

\subsection{The least stable wavelength}

\begin{figure*}
\includegraphics[scale=.98]{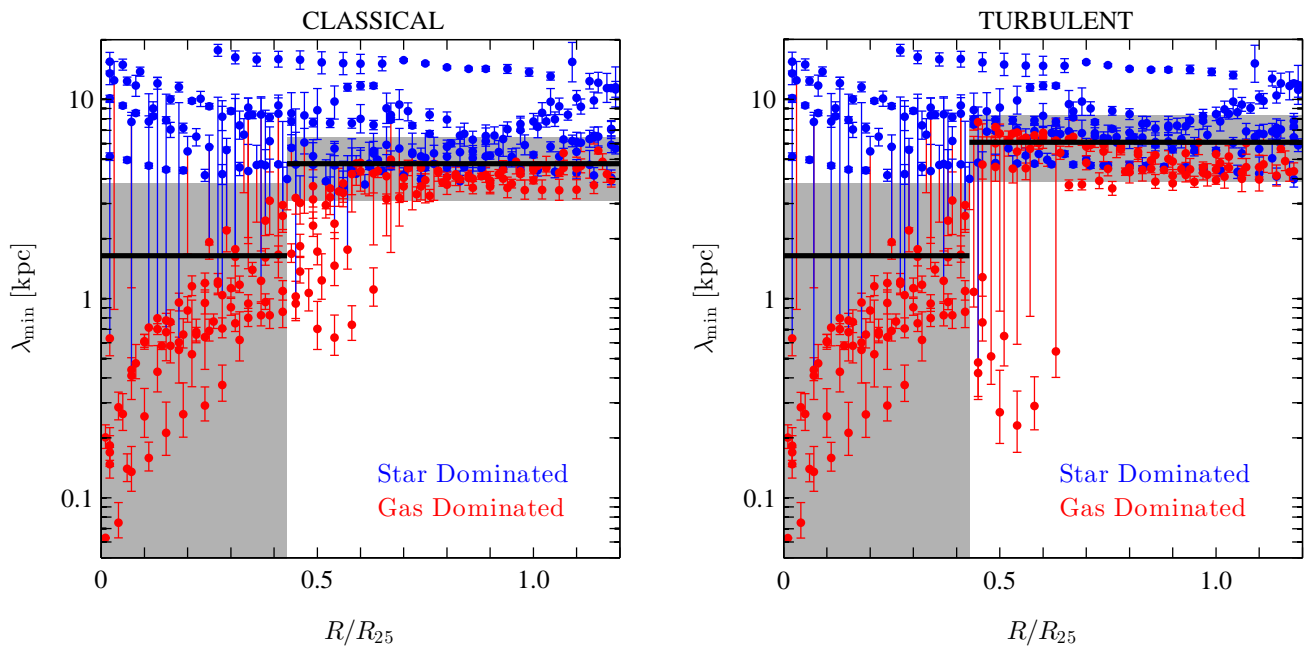}
\caption{Radial profile of the least stable wavelength,
  $\lambda_{\mathrm{min}}(R)$, for THINGS spirals (Model 1): neglecting gas
  turbulence (left), and taking into account H\,\textsc{i} turbulence for
  $R>0.43\;R_{25}$ (right).  For each measurement, colour-coding indicates
  whether gas ($Q_{\mathrm{g}0}<Q_{\star}$) or stars
  ($Q_{\star}<Q_{\mathrm{g}0}$) dominate the stability level.  Thick black
  lines and dark grey shading indicate the median and $1\sigma$ scatter of
  $\lambda_{\mathrm{min}}$ in the two radial ranges.}
\end{figure*}

Fig.\ 7 shows radial profiles of the least stable wavelength, $\lambda_{\mathrm{min}} = \lambda_{\mathrm{min}}(R)$, for our sample. On the left we neglect gas turbulence, whereas on the right we consider turbulent H\,\textsc{i} for $R > 0.43 \; R_{25}$. Colour-coding indicates the component that dominates gravitational instability. As before, the median and $1\sigma$ scatter are indicated separately for small and large radii.

For $R \leq 0.43 \; R_{25}$, there is a clear gap between gas- and star-dominated points (the median value of $\lambda_\mathrm{min}$ is $\lambda_{\mathrm{min,g}} \approx 0.7 \; \mathrm{kpc}$ and $\lambda_{\mathrm{min},\star} \approx 8.2 \; \mathrm{kpc}$ in the two cases). So the gas-dominated points are characterized by much smaller values of $\lambda_{\mathrm{min}}$. The discrepancy is less significant for $R > 0.43 \; R_{25}$, apart from a few measurements close to $R = 0.43 \; R_{25}$ ($\lambda_{\mathrm{min,g}} \approx 3.9 \; \mathrm{kpc}$ and $\lambda_{\mathrm{min},\star} \approx 6.3 \; \mathrm{kpc}$).

Introducing a turbulent gas component for $R > 0.43 \; R_{25}$ causes a significant increase in $\lambda_{\mathrm{min}}$. For $a = b = \frac{1}{3}$, the median of $\lambda_{\mathrm{min}}$ increases by $28\%$ and the $1\sigma$ scatter by $34\%$. For $a = b = \frac{1}{2}$, the median increases by $41\%$ and the increase in $1\sigma$ scatter is again $34\%$. This suggests a tendency of turbulence to boost the least stable wavelength. As for $Q_{\mathrm{eff}}$, the magnitude of this effect depends on the non-turbulent value of $\lambda_{\mathrm{min}}$. There is a small number of gas-dominated measurements for which the least stable wavelength decreases, but these have large uncertainties.

Why does turbulence affect $\lambda_{\mathrm{eff}}$ more than $Q_{\mathrm{eff}}$? The answer is twofold. First, for a purely gaseous disc $\lambda_{\mathrm{min}}$ increases markedly with $Q_{\mathrm{eff}}$ (Romeo et al. 2010), so that any change in $Q_{\mathrm{eff}}$ will be amplified in $\lambda_{\mathrm{min}}$. Second, as stars are taken into account, gas-dominated points can enter the star-dominated regime, where $\lambda_{\mathrm{min}}$ is much larger (see Sect.\ 2.4). Both effects depend on the power-law slopes $a$ and $b$. They sum up and drive $\lambda_{\mathrm{min}}$ to significantly larger values.

\subsection{Robustness of the results}

Modelling the gas disc as a single component with an intermediate value of
$\sigma_{\mathrm{g}}$ is not the best that can be done.  Here we will no
longer follow this traditional approach.  We will model the gas disc as made
of two components, each with the more representative value of
$\sigma_{\mathrm{g}}$.  A simple way to do it is to treat the inner part of
the disc as $\mathrm{H}_{2}$ dominated and the outer part as H\,\textsc{i}
dominated.  We then set $\sigma_{\mathrm{g}}=6\;\mbox{km\,s}^{-1}$ for
$R\leq0.43\;R_{25}$ and $\sigma_{\mathrm{g}}=11\;\mbox{km\,s}^{-1}$ for
$R>0.43\;R_{25}$ (cf.\ introductory part of Sect.\ 3).

Besides $\sigma_{\mathrm{g}}$, there is another quantity that deserves
particular attention: the stellar radial velocity dispersion, which we now
denote with $\sigma_{R\star}$.  Leroy et al.\ (2008) inferred
$\sigma_{R\star}$ from the vertical velocity dispersion, $\sigma_{z\star}$,
assuming that $(\sigma_{z}/\sigma_{R})_{\star}=0.6$.  In turn,
$\sigma_{z\star}$ was inferred from the stellar exponential scale height,
$H_{\star}$, using the relation $H_{\star}=\sigma_{z\star}^{2}/2\pi
G\Sigma_{\star}$.  Gerssen \& Shapiro Griffin (2012) showed that
$(\sigma_{z}/\sigma_{R})_{\star}$ decreases markedly from early- to late-type
spirals.  The average Hubble stage of THINGS spirals is $\langle T\rangle=4$,
which corresponds to galaxy type Sbc (the mean and the median of $T$ are
equal).  The best-fitting model of Gerssen \& Shapiro Griffin (2012) then
yields $(\sigma_{z}/\sigma_{R})_{\star}=0.5$ (see their fig.\ 4).  Concerning
$H_{\star}$, the relation used by Leroy et al.\ (2008) is not correct.  It is
the total surface density in the disc that determines the stellar exponential
scale height: $H_{\star}=\sigma_{z\star}^{2}/2\pi G\Sigma_{\mathrm{tot}}$,
where $\Sigma_{\mathrm{tot}}=\Sigma_{\star}+\Sigma_{\mathrm{g}}$ (Bahcall \&
Casertano 1984; Romeo 1992).  In view of these facts, we set
$(\sigma_{z}/\sigma_{R})_{\star}=0.5$ and use the correct relation for
$H_{\star}$.

Finally, we implement gas turbulence as in Model 1, i.e.\ only for
$R>0.43\;R_{25}$, where the disc is H\,\textsc{i} dominated.  This is simply
because the constant spatial resolution of 800 pc used by Leroy et
al.\ (2008) is too coarse to probe the range of scales affected by
$\mathrm{H}_{2}$ turbulence [see Sect.\ 2.4, case (i)].  Hereafter we will
refer to the model described above as \emph{Model 2}.

\begin{table*}
\begin{minipage}{135mm}
\caption{Stability characteristics of THINGS spirals for Models 1 and 2.}
\begin{center}
\begin{tabular}{c c c c r r r r c r}
\hline

    Model
  & Radial Range
  & $a$ & $b$
  & $f_2$\textsuperscript{a}
  & $f_{2,\mathrm{g}}$\textsuperscript{b}
  & $f_\mathrm{u}$\textsuperscript{c}
  & $f_\mathrm{g}$\textsuperscript{d}
  & $Q_\mathrm{eff}$\textsuperscript{e}
  & $\lambda_\mathrm{min}$\textsuperscript{f} $\left[ \mathrm{kpc} \right]$\\

\hline
  
    $1$
  & $R \leq 0.43 \, R_{25}$
  & $0$ & $0$ 
  & $61\%$
  & $68\%$
  & $13\%$
  & $56\%$
  & $1.67 \pm 0.68$
  & $1.64 \pm 2.17$ \\

  & $R > 0.43 \, R_{25}$
  & $0$ & $0$
  & $4\%$
  & $100\%$
  & $4\%$
  & $39\%$
  & $1.50 \pm 0.46$
  & $4.76 \pm 1.67$ \\

  &
  & $1/3$ & $1/3$
  & $22\%$
  & $41\%$
  & $3\%$
  & $39\%$
  & $1.55 \pm 0.53$
  & $6.10 \pm 2.24$ \\

  &
  & $1/2$ & $1/2$
  & $52\%$
  & $9\%$
  & $3\%$
  & $39\%$
  & $1.59 \pm 0.58$
  & $6.70 \pm 2.23$ \\

  \hline

    $2$
  & $R \leq 0.43 \, R_{25}$
  & $0$ & $0$ 
  & $73\%$
  & $76\%$
  & $25\%$
  & $77\%$
  & $1.50 \pm 0.91$
  & $0.67 \pm 0.62$ \\

  & $R > 0.43 \, R_{25}$
  & $0$ & $0$
  & $4\%$
  & $100\%$
  & $0.5\%$
  & $39\%$
  & $1.99 \pm 0.57$
  & $9.32 \pm 3.40$ \\

  &
  & $1/3$ & $1/3$
  & $19\%$
  & $48\%$
  & $0.5\%$
  & $39\%$
  & $2.09 \pm 0.66$
  & $11.10 \pm 2.78$ \\

  &
  & $1/2$ & $1/2$
  & $52\%$
  & $5\%$
  & $0.5\%$
  & $39\%$
  & $2.14 \pm 0.72$
  & $12.00 \pm 2.43$ \\

\hline

\end{tabular}
\end{center}
\raggedright
\textsuperscript{a} Fraction of data that fall within the two-phase region.\\
\textsuperscript{b} Fraction of the data points in \textsuperscript{a} that populate the gaseous phase.\\
\textsuperscript{c} Fraction of data such that $Q_\mathrm{eff} < 1$.\\
\textsuperscript{d} Fraction of data such that $Q_{\mathrm{g}0} < Q_\star$.\\
\textsuperscript{e} Median and $1\sigma$ scatter of $Q_\mathrm{eff}$.\\
\textsuperscript{f} Median and $1\sigma$ scatter of $\lambda_\mathrm{min}$.\\
\end{minipage}
\end{table*}

Table 1 summarizes the dynamical differences between Model 2 and Model 1. On the whole, the stability diagnostics are moderately affected by the model. The most sensitive diagnostic is $\lambda_\mathrm{min}$, which differs by a factor of 2--3. $Q_\mathrm{eff}$ is more robust, with a difference well below a factor of 2. In Model 2, both $\lambda_\mathrm{min}$ and $Q_\mathrm{eff}$ are smaller for $R \leq 0.43 \; R_{25}$ and larger for $R > 0.43 \; R_{25}$.

Despite these differences, the effect of turbulence is comparable in the two models. For $R \leq 0.43 \; R_{25}$, $f_2$ and $f_{2,\mathrm{g}}$ are slightly larger in Model 2. So H$_2$ is more decoupled from stars and slightly more dominant. For $R > 0.43 \; R_{25}$, $f_2$ is almost identical in the two models, irrespective of the value of $a=b$. Turbulence increases the median value of $Q_\mathrm{eff}$ by less than 10\% in both models. In contrast, the median value of $\lambda_\mathrm{min}$ increases by 20--30\% in Model 2, i.e. less than in Model 1. Summarizing, the effect of H\textsc{i} turbulence in Model 2 is only slightly weaker than in Model 1. This points to the robustness of our results.

\section{DISCUSSION}

Our results cannot be directly compared with those of Shadmehri \& Khajenabi
(2012), hereafter SK12.  This is partly because of the wider scope of our
paper, which embraces a brand-new application to THINGS spirals, and because
most of the analysis carried out by SK12 cannot be easily interpreted.

SK12 analysed five stability regimes of gas turbulence: $a>1$ and
$b<\frac{1}{2}\,(1+a)$; $a=1$ and $b\neq1$; and Regimes A--C.  The first
regime corresponds to a fractal dimension $D=a+2$ higher than 3, and is
therefore beyond the natural range of $a$ (see fig.\ 1 and sect.\ 3 of Romeo
et al.\ 2010).  In the second regime, the volume density is scale-independent
($D=3$), so the medium is incompressible and hence subsonic.  Cold
interstellar gas is instead dominated by compressible structures and
supersonic motions.  Therefore even this regime is of marginal interest (see
again fig.\ 1 and sect.\ 3 of Romeo et al.\ 2010).  Regimes B and C are
populated by $\mathrm{H}_{2}$ turbulence, which manifests itself at scales
less than $L_{\mathrm{H2}}\sim100\;\mbox{pc}$.  In turn, $L_{\mathrm{H2}}$ is
one order of magnitude smaller than the characteristic scale of stellar
instabilities.  Therefore stars play a negligible role in these stability
regimes [see Sect.\ 2.4, case (i)].  SK12 reached the opposite conclusion.
But this is because they assumed Larson-type scaling relations even at kpc
scales, disregarding the type of turbulence associated with such regimes.
Regime A is populated by both $\mathrm{H}_{2}$ and H\,\textsc{i} turbulence.
While the $\mathrm{H}_{2}$ case raises the same issue as Regimes B and C, the
H\,\textsc{i} case is conceptually simpler.  H\,\textsc{i} turbulence
manifests itself at all scales of galactic interest, so stars can play a
significant role in this stability regime [see Sect.\ 2.4, case (ii)].  SK12
reached a similar conclusion.  However, even in this case, their approach is
different from ours.  They chose $a$, $b$ and $\mathcal{L}_{0}$ so as to
sample Regime A, and studied the dispersion relation numerically.  We have
instead examined the whole regime analytically (see Sect.\ 2.3.1).  We have
then chosen observationally motivated values of $a$, $b$ and
$\mathcal{L}_{0}$, and analysed the onset of gravitational instability in the
disc (see in particular Sects 2.4.1 and 2.4.2).

In conclusion, there is a fundamental difference between our analysis and
that of SK12.  Our analysis takes into account the astrophysical relevance of
the various stability regimes, as well as the tight constraints imposed by
observations of ISM turbulence in the Milky Way and nearby galaxies.  These
are important aspects of the problem, which are missing from their analysis.

\section{CONCLUSIONS}

Our analysis of THINGS spirals shows that H\,\textsc{i} turbulence has a
triple effect on the outer regions of galactic discs: (i) it weakens the
coupling between gas and stars in the development of disc instabilities, (ii)
it makes the disc more prone to star-dominated than gas-dominated
instabilities, and (iii) it typically increases the least stable wavelength
by 20--40\% (the steeper the H\,\textsc{i} scaling relations, the larger the
effect).  This is in contrast to the typical 3--8\% increase predicted for
the effective $Q$ parameter.  The effect of H\,\textsc{i} turbulence is in a
sense complementary to the effect of disc thickness.  In fact, disc thickness
increases the effective $Q$ parameter by 20--50\% (Romeo \& Wiegert 2011) but
hardly changes the least stable wavelength (Romeo 1992, 1994) or the
condition for star-gas decoupling (Romeo \& Wiegert 2011).

Our analysis of THINGS spirals also suggests that $\mathrm{H}_{2}$ turbulence
has a significant effect on the inner regions of galactic discs.  For
$R\la0.4\,R_{25}$, i.e.\ where $\mathrm{H}_{2}$ dominates over H\,\textsc{i},
60--70\% of the data fulfil the condition for star-gas decoupling and
70--80\% of these points represent gas-dominated stability regimes.  In such
cases, the onset of gravitational instability is controlled by
$\mathrm{H}_{2}$.  Turbulence is expected to play an important role in this
process at scales smaller than about 100 pc (see Sect.\ 2.4).  If $a=0$ and
$b=\frac{1}{2}$, then $\mathrm{H}_{2}$ turbulence drives the disc to a regime
of transition between instability at small scales and stability \`{a} la
Toomre, as was first pointed out by Romeo et al.\ (2010) in the case of
one-component turbulent discs.  Since this is a regime of transition, even
small deviations from the standard $\mathrm{H}_{2}$ scaling laws ($a=0$ and
$b=\frac{1}{2}$) can have a strong impact on the gravitational instability of
the disc.  This is true even when the mass densities of H\,\textsc{i} and
$\mathrm{H}_{2}$ are comparable, since small-scale instabilities are more
actively controlled by $\mathrm{H}_{2}$ (see Sect.\ 2.3).

\section*{ACKNOWLEDGMENTS}

We are very grateful to Oscar Agertz, Christoph Federrath, Mathieu Puech and
Joachim Wiegert for useful discussions.  We are also grateful to an anonymous
referee for constructive comments and suggestions, and for encouraging future
work on the topic.  ABR thanks the warm hospitality of both the Department of
Physics at the University of Gothenburg and the Department of Fundamental
Physics at Chalmers.

\bsp

\label{lastpage}

\end{document}